\documentclass[conference]{IEEEtran}

\IEEEoverridecommandlockouts

\usepackage{cite}
\usepackage{amsmath,amssymb,amsfonts}
\usepackage{algorithmic}
\usepackage{graphicx}
\usepackage{textcomp}
\usepackage{xcolor}
\def\BibTeX{{\rm B\kern-.05em{\sc i\kern-.025em b}\kern-.08em
    T\kern-.1667em\lower.7ex\hbox{E}\kern-.125emX}}

\usepackage{fancyhdr}
\usepackage{hyperref}
\usepackage{url}

\usepackage{multirow}

\usepackage{stackengine}

\usepackage{diagbox}

\usepackage{courier}

\usepackage{subcaption}

\usepackage{tabularx}

\usepackage{adjustbox}

\usepackage{array, makecell}

\usepackage{lipsum}

\usepackage{xcolor}

\usepackage{accents}

\usepackage{dsfont}

\usepackage[switch]{lineno}

\usepackage{makecell}

\let\vec\mathbf

\begin{document}

\title{Motor Imagery Decoding Using Ensemble Curriculum Learning and Collaborative Training
\thanks{The work of Georgios Zoumpourlis was supported by QMUL Principal's Studentship.}
}

\author{\IEEEauthorblockN{Georgios Zoumpourlis}
\IEEEauthorblockA{\textit{School of Electronic Engineering and Computer Science} \\
\textit{Queen Mary University of London}\\
g.zoumpourlis@qmul.ac.uk}
\and
\IEEEauthorblockN{Ioannis Patras}
\IEEEauthorblockA{\textit{School of Electronic Engineering and Computer Science} \\
\textit{Queen Mary University of London}\\
i.patras@qmul.ac.uk}
}


\maketitle

\thispagestyle{fancy}
\IEEEpubidadjcol


\begin{abstract}
In this work, we study the problem of cross-subject motor imagery (MI) decoding from electroencephalography (EEG) data. Multi-subject EEG datasets present several kinds of domain shifts due to various inter-individual differences (e.g. brain anatomy, personality and cognitive profile). These domain shifts render multi-subject training a challenging task and also impede robust cross-subject generalization. Inspired by the importance of domain generalization techniques for tackling such issues, we propose a  two-stage model ensemble architecture built with multiple feature extractors (first stage) and a shared classifier (second stage), which we train end-to-end with two novel loss terms. The first loss applies curriculum learning, forcing each feature extractor to specialize to a subset of the training subjects and promoting feature diversity. The second loss is an intra-ensemble distillation objective that allows collaborative exchange of knowledge between the models of the ensemble. We compare our method against several state-of-the-art techniques, conducting subject-independent experiments on two large MI datasets, namely PhysioNet and OpenBMI. Our algorithm outperforms all of the methods in both 5-fold cross-validation and leave-one-subject-out evaluation settings, using a substantially lower number of trainable parameters. We demonstrate that our model ensembling approach combining the powers of curriculum learning and collaborative training, leads to high learning capacity and robust performance. Our work addresses the issue of domain shifts in multi-subject EEG datasets, paving the way for calibration-free brain-computer interfaces. We make our code publicly available at: \url{https://github.com/gzoumpourlis/Ensemble-MI}.
\end{abstract}


\section{Introduction}
\label{sec:01_intro}

Brain-Computer Interfaces (BCIs)~\cite{yadav_2020_bcireview} are communication systems that enable human users to interact with computers, robotic limbs or wheelchairs, translating brain activity into commands. BCIs have a wide spectrum of applications, including post-stroke rehabilitation of limb motor impairments~\cite{baniqued_2021_rehabreview}, character typing through visual spellers~\cite{xu_2020_speller} and interactive image generation~\cite{spape_2021_attractive}. The operation of BCI systems leverages neuroimaging techniques to collect brain signals, with the most prevalent one being electroencephalography (EEG)~\cite{berger_1929_eeg}. Advancing EEG-based BCIs towards out-of-the-lab settings, requires equipping them with machine learning models that have robust cross-subject generalization.

With the advent of deep learning (DL), significant steps have been made in the exploration of training methodologies and model architectures that can accommodate learning from EEG datasets with increasingly large number of participants. Inter-subject variability is one of the biggest challenges for EEG-based BCIs, referring to the existence of differences in the characteristics of EEG signals acquired from different individuals (i.e., there are different data distributions for each subject)~\cite{ma_2019_reducing}. In the literature, often the data of each individual are considered as a separate domain~\cite{kostas_2020_thinker}, hence inter-subject differences are treated as domain shifts.

In this work, we consider the problem of EEG-based motor imagery (MI) decoding in subject-independent settings. Motor imagery is a well-known paradigm for BCIs, involving the imagination of motor acts, without overt motor execution or muscle activation~\cite{lotze_2006_volition}. The usage of the MI paradigm is based on the phenomenon of sensorimotor rhythms~\cite{yuan_2014_reviewsmr}, i.e. rhythmic oscillations over the sensorimotor cortex that are modulated during motor imagery. There are several sources of variation that lead to domain shifts in cross-subject MI decoding problems, such as personality type, cognitive profile, neurophysiological predictors, brain anatomy and familiarity with BCI technology~\cite{jeunet_2014_impact, jeunet_2015_predicting}. These variations in turn lead to spatial, spectral and temporal differences in the manifestation of sensorimotor rhythms across individuals~\cite{rimbert_2022_erders}. These differences are the sources of domain shifts that we aim to overcome in order to obtain robust performance in subject-independent settings.

Domain generalization has been explored as a learning paradigm for building convolutional neural networks (CNNs) with strong cross-subject accuracy. A category of domain generalization techniques~\cite{wang_2022_dgsurvey} that has been successfully applied on EEG-based problems~\cite{bakas_2022_beetl, reuben_2020_ensemblingcrowd} is ensemble learning~\cite{zhou_2021_dael}. A resulting property of ensembling is the emergence of diverse feature representations, leading to better generalization. However, existing ensembling works achieve this diversity at the cost of increased computational complexity~\cite{du_2022_ienet} and lengthy model selection procedures~\cite{dolzhikova_2021_ensemble} that cannot be implemented in a single end-to-end trainable pipeline. Another line of works explore multi-branch architectures, assigning a separate branch per training subject~\cite{wei_2021_scsn} or per EEG frequency band~\cite{ma_2022_inceptionnetmultibranch}. Such approaches are compromised by the fact that they focus on individual aspects of the feature extraction, model training or model selection processes. 

We argue that further unleashing the potential of ensembling methods for representation learning on neural signals, requires to jointly consider the design of network architectures and training objectives. We frame our approach as a model ensembling method combined with: (i) a curriculum learning strategy to promote the diversity on individual models and (ii) a collaborative training scheme to exchange knowledge between the models through a distillation loss. We design a training curriculum, such that each model of the ensemble is trained on \textit{all} the source domains (i.e. training subjects), yet progressively specializes to a specific subset of subjects. This leads each model to capture patterns that are mostly specific to the EEG signal characteristics of a subset of training subjects, rather than the entire training set. Training our architecture under such a curriculum, equips it with strong generalization capabilities, by covering a wide range of patterns through several models that act as diverse feature extractors. To regulate the trade-off between diversity and generalization~\cite{bian_2021_diversity}, we introduce an intra-ensemble distillation loss that pushes the predictions of each individual model close to the average of the predictions of all the other models, thereby controlling the diversity between the models of the ensemble. In essence, our collaborative training scheme leads to distillation of knowledge \textit{across} models, working complementary with the curriculum that is applied \textit{within} each model. The balance between diversity and generalization is controlled through a hyperparameter that weighs the contribution of the distillation loss to the total loss.

Our contributions are the following:
\begin{itemize}
    
    \item We propose a model ensembling architecture which we pair with a novel curriculum learning scheme. Our curriculum promotes diversity on the models of the ensemble, driving each model to specialize to a different subset of training subjects. To our knowledge, curriculum learning has not been previously explored for cross-subject MI decoding.
    
    \item We propose an auxiliary intra-ensemble distillation loss, allowing the exchange of knowledge between the individual models of the ensemble. This balances the diversity-generalization trade-off, leading to further performance improvement. Our work is the first to propose a pseudo-labelling scheme for EEG-based knowledge distillation.
    
    \item We conduct our experimental analysis on two large motor imagery datasets (PhysioNet~\cite{goldberger_2000_physionet} and OpenBMI~\cite{lee_2019_openbmi}) totalling more than 150 subjects. We compare our method against four state-of-the-art techniques, namely TIDNet~\cite{kostas_2020_thinker}, EEGSym~\cite{perez_2022_eegsym}, MIN2Net~\cite{autthasan_2021_min2net} and ATL~\cite{zhang_2021_adaptivetl}, showing superior results.
    
    \item We make our code publicly available\footnote{\url{https://github.com/gzoumpourlis/Ensemble-MI}} to support reproducibility.
    
\end{itemize}

The rest of the manuscript is organized as follows. In Section~\ref{sec:02_related_work} we outline relevant previous works, while in Section~\ref{sec:03_proposed_method} we describe our proposed method. In Section~\ref{sec:04_experimental_results} we present the results of our experimental analyses and ablation studies. Lastly, in Section~\ref{sec:05_conclusion} we conclude the manuscript.

\section{Related Work}
\label{sec:02_related_work}

In this Section, we present an overview of the related work on the topics of domain generalization, ensemble learning and feature diversity.

\textbf{Domain generalization:} Several works building subject-independent models for EEG data (which by nature is a domain generalization problem), do not explicitly take care of inter-subject variability~\cite{autthasan_2021_min2net, zhu_2022_bcimodels, perez_2022_eegsym}. Such methods adopt approaches based on Empirical Risk Minimization (ERM)~\cite{vapnik_1998_statistical} that simply minimize the training loss over all source domains (i.e. training subjects). Other methods that have been occasionaly used for multi-subject EEG training are Euclidean/Riemannian Alignment (EA/RA)~\cite{he_2019_aligneucl, kostas_2020_thinker, xu_2020_cross}, which are powerful baselines for learning domain-invariant representations. In our work, we leverage the benefits of RA, along with our proposed curriculum learning and intra-ensemble distillation techniques.

\textbf{Ensemble learning:} A successful example of model ensembling using CNNs is the work of~\cite{bakas_2022_beetl}, where a $k$-fold cross validation process results in $k$ trained models, with each model trained on data from all the available training subjects.~\cite{reuben_2020_ensemblingcrowd} leverage the power of available crowdsourced algorithms for an EEG-based seizure prediction competition~\cite{kuhlmann_2018_epilepsyecosystem}, exploring the possibility of obtaining performance improvements by combining them through model ensembling. The model ensemble proposed by~\cite{dolzhikova_2021_ensemble} requires multiple hyperparameter tuning runs to train each base model. In IENet~\cite{du_2022_ienet}, an ensemble of models with convolutional layers of varying kernel length across multiple scales is utilized to extract rich feature representations. However, each base model of IENet has more than seventy convolutional layers, bringing into question the practicality and interpretability of the proposed architecture.

\textbf{Feature diversity:} One of the key properties of ensemble learning, is the emergence of diverse feature representations across the individual base models of ensembles. Feature diversity can also be obtained through alternative techniques which do not fall within the category of ensemble learning, as they explore ways to obtain diverse features through a single model. ~\cite{ma_2022_inceptionnetmultibranch} propose a multi-branch network architecture where the input EEG signal is divided in four frequency bands, with a dedicated branch for each band.~\cite{altuwaijri_2022_eegnetmultibranch} introduce a multi-branch network based on EEGNet~\cite{lawhern_2018_eegnet}, where each branch contains a different number of temporal filters, as well as a different temporal filter length.~\cite{wei_2021_scsn} propose a multi-branch Separate-Common-Separate Network (SCSN) to tackle the issue of negative transfer learning. Negative learning can appear when training subject-agnostic feature extractors, i.e. when all the layers of a single model are trained on all the training subjects. As a remedy to this, SCSN has a separate feature extractor for each training subject. However we claim that such an approach leads to non-optimal solutions, as training subject-specific layers compromises their generalization capability. We propose a model ensembling approach that differs from these two scenarios (i.e. subject-specific or subject-agnostic layers), yet combines the best of both worlds. In contrast to SCSN, we train multiple feature extraction models on all training subjects, yet we guide each individual feature extractor to specialize on a subset of multiple subjects.

\section{Proposed Method}
\label{sec:03_proposed_method}

\begin{figure*}[!h]
\begin{center}
  \includegraphics[width=1.0\linewidth]{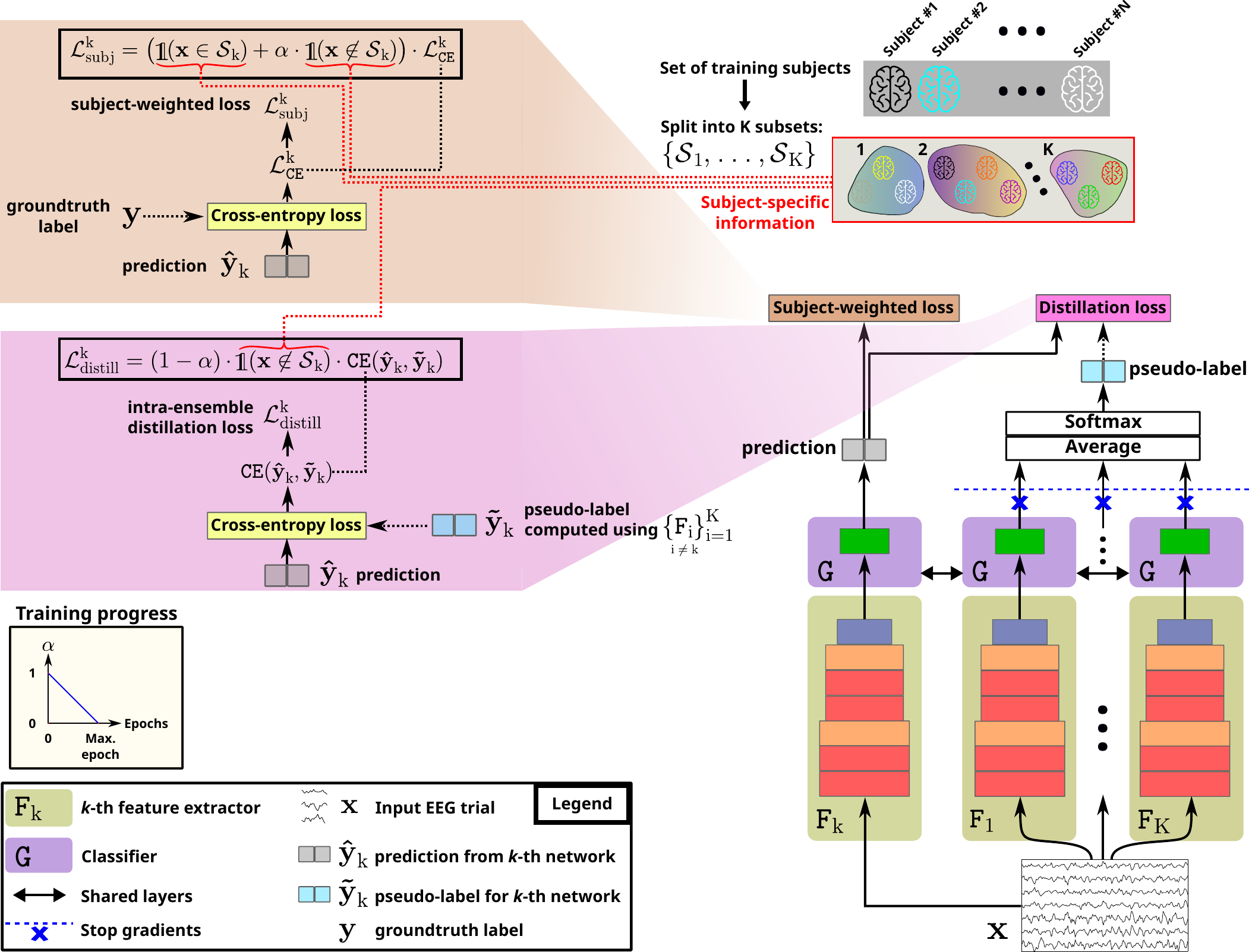}
\end{center}
\caption{Our proposed architecture has $\mathrm{K}$ first stage feature extractors and a shared classifier in the second stage. Each feature extractor is trained: i) on an ensemble curriculum learning objective ($\mathcal{L}_{\mathrm{subj}}$) and ii) on a knowledge distillation objective ($\mathcal{L}_{\mathrm{distill}}$).}
\label{fig:arch_losses}
\end{figure*}

In this section we describe the proposed methodology, which consists of a model ensemble architecture, a curriculum training scheme and an intra-ensemble distillation loss. We provide an overview of the training pipeline for our proposed architecture in Fig.~\ref{fig:arch_losses} and present its individual components in the following subsections. Specifically, we begin by explaining our ensemble architecture in Subsection~\ref{subsec:architecture}. Then, we introduce the first loss term that materializes our curriculum learning scheme in Subsection~\ref{subsec:ens_curr_learning}, as well as the second loss term that enables collaborative training across the models of the ensemble in Subsection~\ref{subsec:ens_distillation}.

\subsection{Architecture}
\label{subsec:architecture}

\textbf{Single model}: In this work, we use the well-established EEGNet~\cite{lawhern_2018_eegnet} architecture as our strong single-model baseline. The selection of EEGNet is justified from the fact that it achieves compelling performance, with a reasonably small number of trainable parameters and a simple network design (e.g. without streams of varying kernel lengths, or band-wise processing streams). In the task of MI decoding, the time-series signals $\vec{x} \in \mathbb{R}^{\mathrm{C} \times \mathrm{T}}$ of an EEG trial with $\mathrm{C}$ electrodes and $\mathrm{T}$ samples in the temporal dimension, are fed as input to EEGNet. The class-wise scores $\vec{\hat{y}} \in \mathbb{R}^{\mathrm{N_C}}$ (where $\mathrm{N_C}$ is the number of classes) are obtained as output, while the groundtruth label $\vec{y} \in \mathbb{R}^{\mathrm{N_C}}$ is represented in the form of a one-hot vector. Thus, in the case of EEGNet the output scores are computed as $\vec{\hat{y}} = \texttt{EEGNet}(\vec{x})$

and the network is optimized by minimizing the cross-entropy (\texttt{CE}) loss $\mathcal{L}_{\texttt{CE}} = \texttt{CE}(\vec{\hat{y}}, \vec{y})$, given by $\texttt{CE}(\vec{\hat{y}}, \vec{y}) = -\sum_{\mathrm{i}=1}^{\mathrm{N_C}} y_{\mathrm{i}} \log\Bigl( \texttt{softmax}( \hat{y}_{\mathrm{i}} ) \Bigr)$

where $y_{\mathrm{i}}$ and $\hat{y}_{\mathrm{i}}$ are the $i$-th elements of $\vec{y}$ and $\vec{\hat{y}}$ respectively.

\textbf{Model ensemble}: Our model ensemble architecture (shown in Fig.~\ref{fig:arch_losses}) consists of two stages and uses EEGNet as its elementary component. The first stage contains multiple models in parallel, with all models having exactly the same architecture design. These models act as feature extractors on an input sample, with each model producing a feature vector. We use $\texttt{F}_{\texttt{k}}(\cdot)$ and $\vec{f_{\mathrm{k}}}$ to denote the $k$-th feature extractor and its output feature vector. The output feature vectors from the first stage, are computed as $[ \vec{f_{\mathrm{1}}}, \vec{f_{\mathrm{2}}}, \ldots, \vec{f_{\mathrm{K}}} ] = [ \texttt{F}_{\texttt{1}}(\vec{x}), \texttt{F}_{\texttt{2}}(\vec{x}), \ldots, \texttt{F}_{\texttt{K}}(\vec{x}) ]$.

The second stage has a single shared classification head $\texttt{G}(\cdot)$, that computes the class-wise prediction scores for each feature vector originating from the first stage. We use $\vec{\hat{y}}_{\mathrm{k}}$ to denote the scores corresponding to the $k$-th feature vector $\vec{f_{\mathrm{k}}}$. The scores are computed as $[ \vec{\hat{y}}_{\mathrm{1}}, \vec{\hat{y}}_{\mathrm{2}}, \ldots, \vec{\hat{y}}_{\mathrm{K}} ] = [ \texttt{G}(\vec{f_{\mathrm{1}}}), \texttt{G}(\vec{f_{\mathrm{2}}}), \ldots, \texttt{G}(\vec{f_{\mathrm{K}}}) ]$.

In the simple scenario where no curriculum learning occurs, this architecture is trained by minimizing the sum of the individual losses for the predictions of each model. The loss $\mathcal{L}_{\texttt{CE}}^{\mathrm{k}}$ for the predictions $\vec{\hat{y}}_{\mathrm{k}}$ of the $k$-th model, and the total loss $\mathcal{L}_{\texttt{CE}}^{\mathrm{total}}$, are computed as $\mathcal{L}_{\texttt{CE}}^{\mathrm{k}} = \texttt{CE}(\vec{\hat{y}}_{\mathrm{k}}, \vec{y})$ and $\mathcal{L}_{\texttt{CE}}^{\mathrm{total}} = \sum_{\mathrm{k}=1}^{\mathrm{K}} \mathcal{L}_{\texttt{CE}}^{\mathrm{k}}$.

In the inference phase, to classify an input sample $\vec{x}$ we fuse the model-wise scores through a simple average operation and obtain a final score vector $\vec{\hat{y}}_{\mathrm{ens}}$ as $\vec{\hat{y}}_{\mathrm{ens}} = \frac{1}{\mathrm{K}} \sum_{\mathrm{k}=1}^{\mathrm{K}} \vec{\hat{y}}_{\mathrm{k}}$. To this end, the described architecture is purely subject-agnostic, having no subject-specific layers in both stages. In the following subsection we propose an ensemble curriculum learning scheme that is applied during training and changes the nature of the first stage layers. Our curriculum provides a strong alternative to the typical subject-agnostic layers, that can be adopted in ensemble learning.

\subsection{Ensemble curriculum learning}
\label{subsec:ens_curr_learning}

Our goal is to make each feature extractor to specialize on a specific subset of subjects. That is, we want to induce \textit{local} (i.e. focused on a subset of the entire training set) feature extraction power to each model in the first stage. Let $\mathbf{\mathcal{D}} = \{ \mathbf{\mathcal{D}}_1, \mathbf{\mathcal{D}}_2, \ldots, \mathbf{\mathcal{D}}_\mathrm{N} \}$ be a dataset with the data of $\mathrm{N}$ subjects, where $\mathbf{\mathcal{D}}_\mathrm{n}$ denotes the sub-dataset containing the trials of the $n$-th subject. For an ensemble with $\mathrm{K}$ models ($\mathrm{K} \geq 2$), we split $\mathbf{\mathcal{D}}$ into $\mathrm{K}$ non-overlapping subsets $\mathbf{\mathcal{S}}$: $\mathbf{\mathcal{D}} = \{ \mathbf{\mathcal{S}}_1, \ldots, \mathbf{\mathcal{S}}_{\mathrm{K}} \}$. We do this splitting process by randomly assigning the sub-dataset of each subject to one of the $\mathrm{K}$ subsets, with a uniform probability for all subsets. Therefore, we have $\bigcup_{\mathrm{k}=1}^{\mathrm{K}} \mathbf{\mathcal{S}}_\mathrm{k} = \mathbf{\mathcal{D}}$ and $\mathbf{\mathcal{S}}_{\mathrm{i}} \cap \mathbf{\mathcal{S}}_{\mathrm{j}} = \emptyset$ for $i \neq j$. Each subset $\mathbf{\mathcal{S}}_{\mathrm{k}}$ corresponds to the $k$-th model and contains the sub-datasets of the subjects on which we drive the $k$-th model to specialize.

To achieve this specialization, we design a \textit{subject-weighted} loss function where we inject subject-specific coefficients to weigh the contribution of each subject to the loss of each model. Considering the subject-weighted loss $\mathcal{L}_{\mathrm{subj}}^{\mathrm{k}}$ that is used to train the $k$-th model, the purpose of the subject-specific coefficients is to linearly decay over epochs the loss contribution of the subjects that \textit{do not} belong to $\mathbf{\mathcal{S}}_{\mathrm{k}}$. Effectively, this makes the $k$-th model to focus more on the subjects of $\mathbf{\mathcal{S}}_{\mathrm{k}}$, which have a non-decaying loss contribution. We scale the contribution of a training sample $\vec{x}$ to the loss $\mathcal{L}_{\mathrm{subj}}^{\mathrm{k}}$ through the coefficient $\beta(\vec{x},\mathrm{k})$. If trial $\vec{x}$ corresponds to a subject that belongs in $\mathbf{\mathcal{S}}_{\mathrm{k}}$ (hence $\vec{x} \in \mathbf{\mathcal{S}}_{\mathrm{k}}$), then we keep $\beta(\vec{x},\mathrm{k})=1$ throughout the whole training process. Otherwise ($\vec{x} \notin \mathbf{\mathcal{S}}_{\mathrm{k}}$), we decay $\beta(\vec{x},\mathrm{k})$ from 1 to 0 while training progresses, that is:
\begin{equation}
 \beta(\vec{x},\mathrm{k}) = \left\{
 \begin{array}{ll}
      1 & , \mathrm{if}~\vec{x} \in \mathbf{\mathcal{S}}_{\mathrm{k}}  \\
      \alpha & , \mathrm{if}~\vec{x} \notin \mathbf{\mathcal{S}}_{\mathrm{k}}  \\
 \end{array} 
 \right.
 ,
\label{eq:coef_beta}
\end{equation}
where $\alpha = 1 - \frac{ \mathrm{epoch} } { \mathrm{N}_{\mathrm{epochs}} } \in \left[0, 1\right]$ represents the progression of training, as $\mathrm{N}_{\mathrm{epochs}}$ is the maximum number of training epochs and $\mathrm{epoch}$ is the current epoch. The loss $\mathcal{L}_{\mathrm{subj}}^{\mathrm{k}}$ of the $k$-th model and the total subject-weighted loss $\mathcal{L}_{\mathrm{subj}}^{\mathrm{total}}$ are computed as $\mathcal{L}_{\mathrm{subj}}^{\mathrm{k}} = \beta( \vec{x},\mathrm{k} ) \cdot \mathcal{L}_{\texttt{CE}}^{\mathrm{k}}$ and $\mathcal{L}_{\mathrm{subj}}^{\mathrm{total}} = \sum_{\mathrm{k}=1}^{\mathrm{K}} \mathcal{L}_{\mathrm{subj}}^{\mathrm{k}}$.

\subsection{Intra-ensemble distillation for collaborative training}
\label{subsec:ens_distillation}

In this subsection we propose a collaborative training scheme which helps to regulate the diversity-generalization trade-off in our model ensemble. In order to classify a sample, we extract its first stage representations, feed them to the shared classifier of the second stage and average the individual scores across models. The diversity between the first stage representations of a sample can make the classifier to compute inconsistent class scores across models. This, in turn, can negatively affect the final prediction scores, as they will be the result of fusing multiple contradicting predictions. We observe that, although feature diversity is a desirable property of our ensemble, it can also have an adverse effect on the generalization capabilities.

To overcome this phenomenon, we introduce a loss term that promotes consistency across the multiple model predictions, in order to improve the performance of the entire ensemble. We design our proposed intra-ensemble distillation loss to operate on the predicted scores of the second stage, instead of operating on the features extracted from the first stage. An overview of our distillation loss is shown in Fig.~\ref{fig:arch_losses}. Considering each prediction $\hat{\vec{y}}_{\mathrm{k}}$ of the $k$-th model, our loss pushes it closer to the softmaxed average of the predictions from all the other models (which is the pseudolabel in our distillation loss). Specifically, we compute the pseudolabel $\tilde{\vec{y}}_{\mathrm{k}}$ for the $k$-th model as $\vec{\tilde{y}}_{\mathrm{k}} = \texttt{softmax}\biggl( \frac{1}{\mathrm{K}-1} \sum_{\mathrm{i=1, i \neq k}}^{\mathrm{K}} \vec{\hat{y}}_{\mathrm{i}} \biggr)$ and minimize the cross-entropy loss between the prediction $\hat{\vec{y}}_{\mathrm{k}}$ and the pseudolabel $\vec{\tilde{y}}_{\mathrm{k}}$. We note that we apply a stop-gradient~\cite{chen_2021_exploring} operation on the pseudolabels, as shown in Fig.~\ref{fig:arch_losses}. We do this to ensure that only the weights of the $k$-th model are updated based on this loss term, while the other models remain unaffected. For the $k$-th model, we opt to \textit{not} apply this loss on the samples of $\mathcal{S}_{\mathrm{k}}$. This is done through a binary mask that zeroes out the distillation loss of these samples. We do so, as our curriculum learning objective ensures that the $k$-th model is sufficiently trained on the samples of $\mathcal{S}_{\mathrm{k}}$ through their groundtruth labels $\vec{y}$.

We note that it is necessary to scale the contribution of the intra-ensemble distillation loss to the total loss of the architecture, in accordance with the progress of training. In the beginning of the training process, the weights of the architecture are randomly initialized. Hence, penalizing the distance of individual model predictions from the derived pseudolabels is not so meaningful in the early epochs. As training proceeds, each feature extractor progressively focuses on a subset of subjects and feature diversity increases. As shown later in the experiments, our distillation loss indirectly controls this emerging feature diversity by bringing closer the class scores computed from various first stage features. We linearly increase the contribution of the distillation loss to the total loss, across training epochs, by multiplying it with the scalar $(1-\alpha)$ that quantifies the training progress.
The distillation loss $\mathcal{L}_{\mathrm{distill}}^{\mathrm{k}}$ of the $k$-th model, and the total distillation loss $\mathcal{L}_{\mathrm{distill}}^{\mathrm{total}}$ are computed as $\mathcal{L}_{\mathrm{distill}}^{\mathrm{k}} =   \mathds{1}(\vec{x} \notin \mathbf{\mathcal{S}_{\mathrm{k}}}) \cdot \texttt{CE}( \vec{\hat{y}}_{\mathrm{k}}, \vec{\tilde{y}}_{\mathrm{k}} ) \cdot (1-\alpha)$ and $\mathcal{L}_{\mathrm{distill}}^{\mathrm{total}} = \sum_{\mathrm{k}=1}^{\mathrm{K}} \mathcal{L}_{\mathrm{distill}}^{\mathrm{k}}$.

We compute the total loss $\mathcal{L}_{\mathrm{total}}$ of our architecture as $\mathcal{L}_{\mathrm{total}} = \lambda_{\mathrm{subj}} \cdot \mathcal{L}_{\mathrm{subj}}^{\mathrm{total}} + \lambda_{\mathrm{distill}} \cdot \mathcal{L}_{\mathrm{distill}}^{\mathrm{total}}$, 

where we empirically set $\lambda_{\mathrm{subj}}=\mathrm{K}$ and $\lambda_{\mathrm{distill}}=0.7$.

\section{Experimental results}
\label{sec:04_experimental_results}

\subsection{Datasets}

We apply our method on the problem of motor imagery decoding and work on two large datasets: PhysioNet~\cite{goldberger_2000_physionet, schalk_2004_physionet} and OpenBMI~\cite{lee_2019_openbmi}. First we provide a brief description of the datasets and then we describe the common signal preprocessing pipeline that is followed for both datasets.

\textbf{PhysioNet dataset}: The dataset contains EEG recordings from $109$ participants, with trials that belong to $4$ classes: left-hand, right-hand and feet imagery, as well as rest. The data are recorded with $64$ EEG electrodes at a sampling frequency of $160$Hz. We choose to work on the two-class problem of classifying left-hand versus right-hand imaginary movements, discarding the data from the other classes. Similarly to other works~\cite{kostas_2020_thinker, barmpas_2023_causal}, we also discard data from $6$ participants (specifically \texttt{S088}, \texttt{S090}, \texttt{S092}, \texttt{S100}, \texttt{S104} and \texttt{S106}) that have inconsistent sampling frequencies or trial lengths. In our experiments we use the signals from all $64$ electrodes.

\textbf{OpenBMI dataset}: The data of OpenBMI correspond to trials of $2$ classes (left-hand and right-hand imagery) collected from the EEG recordings of $54$ participants, with $62$ electrodes at a sampling frequency of $1000$Hz. Each participant has data from two sessions and each session has two runs. The first run of each session is done in an offline manner, i.e. without feedback. The second run is done in an online manner, providing real-time visual feedback to the user. When not otherwise stated, we use a subset of $20$ electrodes and we use the data from the two offline runs (i.e. the first run of the first and second session) for each participant, following the default settings of the MOABB~\cite{jayaram_2018_moabb} benchmark.

\textbf{Preprocessing:} Our preprocessing steps are the following: (i) we remove powerline interferences through notch filtering (ii) we perform bandpass filtering ($4$Hz-$38$Hz) (iii) we resample the signals to $100$Hz and (iv) for each trial, we crop a temporal window of $4$ seconds, starting from its onset event. Upon obtaining the cropped trials, we use the session-wise covariance matrices of the EEG signals and perform Riemannian Alignment on the time-series of each trial, as done in~\cite{zoumpourlis_2022_covmix}.

\subsection{Comparison with other works and baseline}

We compare our proposed method with four state-of-the-art techniques that provide their source code, namely Adaptive Transfer Learning (ATL)~\cite{zhang_2021_adaptivetl}, EEGSym~\cite{perez_2022_eegsym}, TIDNet~\cite{kostas_2020_thinker} and MIN2Net~\cite{autthasan_2021_min2net}. In order to fairly judge the impact of our proposed methodology, we also implement two additional methods: a single model baseline and an ensembling technique using the EEGNet architecture. The baseline method (mentioned as ``EEGNet-Single'') is a single EEGNet model, that serves as a reference for the performance of an EEGNet architecture without ensembling. We implement the ensembling technique by training multiple individual EEGNet models. During inference, we fuse their predictions through a simple averaging operation, to obtain the final prediction. In essence, this method (mentioned as ``EEGNet-Ensemble'') represents a post-training model ensemble.

\textbf{Evaluation settings}: We perform evaluation in two ways: (i) in a 5-fold cross-validation (CV) manner and (ii) in a Leave-One-Subject-Out (LOSO) manner. In the 5-fold CV scenario, we split the subjects of our dataset into 5 disjoint folds and run 5 experiments. In each experiment, we use a different fold as our test set and then assign 3 folds to our training set and the 1 remaining fold to our validation set. In the LOSO scenario for a dataset with $\mathrm{N}$ subjects, we run $\mathrm{N}$ experiments where in the $n$-th experiment we use the data of the $n$-th subject as our test set. In each experiment, we split the remaining $\mathrm{N}-1$ subjects into our training and validation set. Specifically, we assign $80\%$ of these subjects to the training set and the rest $20\%$ to the validation set of the experiment. In both CV and LOSO scenarios, the reported accuracy is the average of the test accuracies across all experiments.

\textbf{Training details}: We train all models (i.e. our proposed method, the single model baseline and the model ensembling method) for $120$ epochs with a batch size of $64$. We use a Stochastic Gradient Descent (SGD) optimizer, setting the momentum to $0.9$ and weight decay to $0.01$. We initialize the learning rate at $0.01$ for the first $60$ epochs and then decrease it to $0.002$ for the remaining $60$ epochs.

\subsection{Results (5-fold cross-validation)}
\label{results:5fold}

In the first part of our experimental analysis we evaluate against methods that provide source code, under a 5-fold cross-validation scenario, without any model adaptation on test data or pretraining on external datasets. We note that these experiments are performed using exactly the same train, validation and test splits, the same trial length and the same number of electrodes for all methods (except for the method of EEGSym that has an architectural requirement of 16 electrodes). Having the same experimental settings enables us to fairly judge the performance of all methods. Table~\ref{tab:5fold_comparison} shows the results of the methods trained on the datasets of PhysioNet and OpenBMI with 5-fold cross-validation. Regarding the reported results of our proposed method, we note that the optimal number of first stage feature extractors $\mathrm{K}$ is inferred from the accuracy on the validation set. Similarly, regarding the reported results of the EEGNet-Ensemble method, the optimal number of individual EEGNet models within an ensemble is chosen based on the validation accuracy.

\textbf{PhysioNet}: Our proposed method presents a substantial boost of $+1.80\%$ over the standard ensemble scenario, reaching an accuracy of $86.36\%$ when we use seven first stage models in our architecture. The model of EEGSym achieves an accuracy of $83.91\%$, using $\sim10\times$ more trainable parameters than the best performing architecture of our method. EEGSym without pretraining on external data, performs worse than both our method and the standard model ensemble. The accuracy of TIDNet ($82.19\%$) is similar to that of our EEGNet-Single baseline model.

\textbf{OpenBMI}: Our proposed method performs superiorly to our baselines, yielding an accuracy of $79.73\%$ when using three first stage networks. The method of MIN2Net has a low performance, with an accuracy of $69.44\%$. Regarding the method of ATL, the accuracy of $77.52\%$ falls behind the results of both our proposed method and our baseline, using $\sim60\times$ more trainable parameters than our proposed method. Our results show that a simple ensemble architecture trained with a curriculum learning scheme and an auxiliary loss can achieve high cross-subject generalization, without any adaptation on test data or complex model architecture.

\begin{table}[!t]
\caption{Performance of various methods on the datasets of PhysioNet and OpenBMI, under 5-fold CV evaluation settings.}
\label{tab:5fold_comparison}
\centering
\begin{tabular}{|c|l|c|c|} 
\hline
\textbf{Dataset} & \textbf{Method}        & \multicolumn{1}{l|}{\textbf{Parameters}} & \multicolumn{1}{l|}{\textbf{Acc. (\%})}  \\ 
\Xhline{1.2pt}
\multirow{4}{*}{\textbf{PhysioNet}} & \textbf{EEGNet-Single} & 2.5K & 82.09 \\
 & \textbf{\makecell{EEGNet-Ensemble\\(8 models)}} & 20.0K & 84.56 \\
\cline{2-4}
 & \textbf{EEGSym~\cite{perez_2022_eegsym}} & 147.8K & 83.91 \\
 & \textbf{TIDNet~\cite{kostas_2020_thinker}} & 694.2K & 82.19 \\
\cline{2-4}
 & \textbf{Ours, $\mathrm{K}$=7} & 15.7K & $\mathbf{86.36}$ \\
\Xhline{1.2pt}
\multirow{4}{*}{\textbf{OpenBMI}} & \textbf{EEGNet-Single} & 1.8K & 78.31 \\
 & \textbf{\makecell{EEGNet-Ensemble\\(8 models)}} & 14.3K & 78.98 \\
\cline{2-4}
 & \textbf{MIN2Net~\cite{autthasan_2021_min2net}} & 37.1K & 69.44 \\
 & \textbf{ATL~\cite{zhang_2021_adaptivetl}} & 278.8K & 77.52 \\
\cline{2-4}
 & \textbf{Ours, $\mathrm{K}$=3} & 4.6K & $\mathbf{79.73}$ \\
\hline
\end{tabular}
\end{table}

\subsection{Results (Leave-One-Subject-Out)}

In this experiment, we compare our method against other state-of-the-art works that report LOSO results on PhysioNet and OpenBMI. We note that we mention the results of these methods as reported in their original works, ensuring that they do not utilise labelled data from the test subjects. The results are shown in Table~\ref{tab:LOSO_comparison}.

\textbf{PhysioNet}: The method of EEGSym achieves state-of-the-art performance reaching an accuracy of $88.56\%$. EEGSym performs transfer learning by pretraining on four external datasets, which proves to be highly valuable. Our proposed method is the best performing model among the works that do not train on external data. We outperform the method of~\cite{barmpas_2023_causal} that trains separate convolutional layers for each training subject. This indicates the existence of more efficient alternatives to complex deep architectures and the incorporation of subject-specific components.

\textbf{OpenBMI}: Our method presents state-of-the-art performance, scoring an accuracy of $85.07\%$ when using all 62 electrodes of OpenBMI and having $\mathrm{K}=4$ first stage models. We outperform all other techniques, including the method of EEGSym that employs pretraining on external data. The geometric deep learning approach of TSMNet~\cite{kobler_2022_spd} presents an accuracy gap of more than $\sim10\%$ from the methods of ATL, EEGSym and our technique. This indicates that
deep architectures operating on covariance matrices of EEG time-series (e.g.~\cite{kobler_2022_spd} and~\cite{kwon_2019_spectrspatmultibranch}), are generally less suitable for cross-subject MI decoding.

\begin{table}[!t]
\caption{Comparison with other state-of-the-art methods on the datasets of PhysioNet and OpenBMI with LOSO evaluation settings. (\textbf{*}) denotes pretraining on external data.}
\label{tab:LOSO_comparison}
\centering
\begin{tabular}{|c|l|c|c|} 
\Xhline{1.2pt}
\textbf{Dataset} & \textbf{Method}        & \multicolumn{1}{l|}{\textbf{Parameters}} & \multicolumn{1}{l|}{\textbf{Accuracy (\%})}  \\ 
\Xhline{1.2pt}
\multirow{3}{*}{\textbf{PhysioNet}} & \textbf{Causal Viewpoint~\cite{barmpas_2023_causal}} & N/A & 83.90 \\
 & \textbf{EEGSym*~\cite{perez_2022_eegsym}} & 147K & $\mathbf{88.56}$ \\
\cline{2-4}
 & \textbf{Ours, $\mathrm{K}$=7} & 15.7K & 85.82 \\
\Xhline{1.2pt}
\multirow{5}{*}{\textbf{OpenBMI}} & \textbf{MIN2Net~\cite{autthasan_2021_min2net}} & 37.1K & 72.03 \\
 & \textbf{TSMNet~\cite{kobler_2022_spd}} & 4.5K & 74.60 \\
 & \textbf{ATL~\cite{zhang_2021_adaptivetl}} & 305K & 84.19 \\
 & \textbf{EEGSym*~\cite{perez_2022_eegsym}} & 147K & 84.72 \\
\cline{2-4}
 & \textbf{Ours, $\mathrm{K}$=4} & 8.7K & $\mathbf{85.07}$ \\
\hline
\end{tabular}
\end{table}

\subsection{Ablation studies}
\label{subsec:ablation}

In our ablation studies we investigate the impact of three components on the performance of our ensemble architecture. The first component is the number of first stage models $\mathrm{K}$ in the architecture. The second component is the loss $\mathcal{L}_{\mathrm{subj}}^{\mathrm{total}}$, that materializes our curriculum learning scheme. The third component is the distillation loss $\mathcal{L}_{\mathrm{distill}}^{\mathrm{total}}$ that enables collaborative training. We concurrently explore the effects of all these component choices, performing a sweep over the hyperparameter $\mathrm{K}$ and trying combinations of our loss terms.

Our first set of experiments (denoted as ``$\mathcal{L}_{\texttt{CE}}$'') corresponds to the scenario of training a model ensemble architecture as described in Sec.~\ref{subsec:architecture}, i.e. without curriculum learning and without our distillation loss. In our second set of experiments (denoted as ``$\mathcal{L}_{\mathrm{subj}}$'') we train our architecture with ensemble curriculum learning, as described in Sec.~\ref{subsec:ens_curr_learning}, i.e. without our distillation loss. In the third experimental run (denoted as ``$\mathcal{L}_{\mathrm{total}}$'') we apply our entire method, training our architecture with the $\mathcal{L}_{\mathrm{total}}$ loss. All experiments are performed with a 5-fold cross-validation setting. The results of our ablation study are shown in Table~\ref{tab:5fold_ablation}.

\begin{table*}[!ht]
\caption{Ablation study on the datasets of PhysioNet and OpenBMI with 5-fold CV evaluation settings. Rows correspond to experiment sets done with different optimization objectives. Columns correspond to the number of first stage models ($\mathrm{K}$) in our architecture.}
\label{tab:5fold_ablation}
\centering
\begin{tabular}{|c|l|l|l|l|l|l|l|} 
\hline
\multirow{2}{*}{\textbf{Dataset}} & \multirow{2}{*}{\textbf{Loss terms}} & \multicolumn{6}{c|}{\textbf{Accuracy (\%)} } \\ 
\cline{3-8} &  & \textbf{K=2} & \textbf{K=3} & \textbf{K=4} & \textbf{K=5} & \textbf{K=6} & \textbf{K=7} \\
\Xhline{1.2pt}
\multirow{3}{*}{\textbf{PhysioNet}} & $\mathcal{L}_{\texttt{CE}}$ & 83.34 & 84.70 & 84.97 & 84.93 & $\mathbf{85.53}$ & 85.34 \\
 & $\mathcal{L}_{\mathrm{subj}}$ & 84.38 & 84.72 & 85.10 & 85.40 & 85.62 & $\mathbf{85.68}$ \\
 & $\mathcal{L}_{\mathrm{total}}$ & 83.76 & 84.78 & 85.02 & 85.48 & 86.04 & $\mathbf{86.36}$ \\
\Xhline{1.2pt}
\multirow{3}{*}{\textbf{OpenBMI}} & $\mathcal{L}_{\texttt{CE}}$ & 79.15 & 79.08 & 78.96 & $\mathbf{79.24}$ & 78.94 & 79.20 \\
 & $\mathcal{L}_{\mathrm{subj}}$ & 79.02 & $\mathbf{79.58}$ & 79.13 & 79.15 & 79.01 & 79.31 \\
 & $\mathcal{L}_{\mathrm{total}}$ & 79.25 & $\mathbf{79.73}$ & 79.53 & 79.46 & 79.10 & 79.66 \\
\hline
\end{tabular}
\end{table*}

\textbf{PhysioNet}: We observe a general trend of increasing accuracy for all our experimental sets, as $\mathrm{K}$ increases up to the value of $7$ (further increasing $\mathrm{K}$ does not yield performance improvements). The only exception is the case where we train our architecture without curriculum learning (i.e. first row in Table~\ref{tab:5fold_ablation}), where the accuracy saturates at $\mathrm{K}=6$. This indicates that training multiple feature extractors by equally fitting them to the entire training set, is a suboptimal approach of training on multiple source domains. Thus, applying our curriculum learning scheme through $\mathcal{L}_{\mathrm{subj}}$ to induce diversity in the feature extractors, is a straightforward step. The results of the second row in Table~\ref{tab:5fold_ablation} verify the positive impact of curriculum learning in our ensemble architecture. When further incorporating our distillation loss in the total optimization objective of our architecture (i.e. third row in Table~\ref{tab:5fold_ablation}), we get additional accuracy boosts in most cases. The beneficial effect of regulating the balance between feature diversity and model generalization through our distillation loss, is higher in the cases of $\mathrm{K}=6$ and $\mathrm{K}=7$ where the accuracy boosts are $+0.42\%$ and $+0.68\%$ respectively. This finding is particularly interesting, showing that the combination of our two loss terms can increase the performance of model ensembles, even when using many feature extraction models. On the contrary, an ensemble architecture trained solely with the standard cross-entropy loss, is more prone to performance saturation.

\textbf{OpenBMI}: The standard ensemble architecture trained without curriculum learning (i.e. first row in Table~\ref{tab:5fold_ablation}) achieves a maximum accuracy of $79.24\%$ when $\mathrm{K}=5$. By using our curriculum learning scheme, we improve the accuracy of our architecture in four out of six cases, achieving a maximum accuracy of $79.58\%$ when $\mathrm{K}=3$. The incorporation of our distillation loss term in the total loss of our architecture (i.e. third row in Table~\ref{tab:5fold_ablation}) provides consistent improvements in all cases. Our best model has an accuracy of $79.73\%$ when $\mathrm{K}=3$, with a boost of $0.65\%$ over its corresponding standard ensemble model.

\section{Conclusion}
\label{sec:05_conclusion}

In this work, we propose a method for cross-subject motor imagery decoding that leverages the combined strengths of model ensembling, curriculum learning and collaborative training. We design an ensemble architecture that is trained end-to-end in a single phase. We show that our curriculum training scheme can induce diversity to the feature extraction models of our architecture, improving its performance over standard ensembling. Our method also benefits from the exchange of knowledge between the models of our ensemble, that occurs through our auxiliary distillation loss. We conduct experiments on the datasets of PhysioNet and OpenBMI, demonstrating state-of-the-art results. Our proposed method outperforms other approaches that try to tackle MI decoding using complex networks~\cite{zhang_2021_adaptivetl, kostas_2020_thinker}, multi-task learning~\cite{autthasan_2021_min2net}, geometric deep learning~\cite{kobler_2022_spd}, subject-specific layers~\cite{barmpas_2023_causal} or pretraining on multiple external datasets~\cite{perez_2022_eegsym}. Our work highlights the importance of feature diversity as a property of model ensembles, paving the way for robust EEG-based domain generalization techniques.


\bibliography{references}
\bibliographystyle{ieee}

\end{document}